\documentclass[11pt,twoside]{article}

\usepackage{asp2006}
\usepackage{epsf}
\usepackage{lscape}
\usepackage{graphicx}
\usepackage{ulem}

\begin{document}
\title{Helium stars as supernova progenitors}   
\author{Roni Waldman \altaffilmark{1}, Lev R. Yungelson \altaffilmark{2} and Zalman Barkat \altaffilmark{1}}   

\altaffiltext{1}{Racah Institute of Physics, The Hebrew University,
Jerusalem 91904, Israel}

\altaffiltext{2}{Institute of Astronomy, Russian Academy of
Sciences, 48
Pyatnitskaya Str, Moscow, Russia}    

\begin{abstract} 
We follow the evolution of helium stars of initial mass $(2.2 -
2.5)\,M_\odot$, and show that they undergo off-center carbon
burning, which leaves behind
${\mathbf \sim 0.01\,M_\odot}$ of unburnt carbon in the inner part of the core.
When the carbon-oxygen core grows to Chandrasekhar mass, the amount
of left-over carbon is sufficient to ignite thermonuclear runaway.
At the moment of explosion, the star will possess an envelope of
several $0.1\,M_{\odot}$, consisting of He, C, and possibly some H,
 perhaps producing a kind of peculiar SN. Based on the results of
 \citet{Waldman2007} for accreting white dwarfs, we expect to get thermonuclear runaway at a
 broad  range of
$\rho_c \approx (1 - 6) \times 10^9 \mathrm{ g\,cm^{-3}}$, depending
on the amount of residual carbon. We verified the feasibility of
this scenario by showing that in a close binary system with initial
masses $(8.5 + 7.7)\,M_{\odot}$\ and initial period of 150 day the
primary produces a helium remnant of $2.3\,M_{\odot}$\ that evolves
further like the model we considered.
\end{abstract}

\section{Introduction}
Type Ia supernovae (SN~Ia) have a relatively small dispersion of
luminosity (the standard deviation in peak blue luminosity is
$\sigma_B \approx 0.4 - 0.5$ mag., \citet{Branch1993ApJL}) and are
being used as distance indicators (``standard candles''), having
especial significance in the effort of determining the cosmological
parameters of our universe.

The long-standing explanation of the SN~Ia phenomenon has been the
explosive burning of degenerate carbon in the core of a
carbon-oxygen white dwarf, which becomes unstable as it grows to
Chandrasekhar mass ($M_{Ch}$) either by accretion from a binary
companion or by a merger of two white dwarfs, following the angular
momentum loss from the system by gravitational wave radiation.
However, theoretical models are still far from self-consistently
producing an evolutionary path towards the progenitor and
reproducing crucial features of the observational data, such as the
composition of the ejecta. For a detailed review of the above see,
e.g., \citet{Leibundgut2000A&ARv, Hillebrandt&Niemeyer2000ARA&A,
Filippenko2005ASSL}.

As well, SN~Ia can not be regarded as perfectly
homogeneous class, since their Hubble
diagram  exhibits scatter larger than the
photometric errors, while spectroscopic and photometric
peculiarities have been noted with increasing frequency in
well-observed SN~Ia \citep[e.g., ][]{Filippenko2005ASSL}.

Therefore, there is an obvious need for progenitor scenarios that
could explain the diversity among SN~Ia. Several explanations have
been suggested, such as variations in the metallicity of the
progenitor, in the carbon to oxygen ratio at its center, or in the
central density at the time of ignition \citep[e.g., ][]{Timmes2003ApJ,
Ropke_etal2006A&A, Lesaffre2006MNRAS}. The variation of the latter
two parameters is expected to result from the variation in the
initial white dwarf mass and in the accretion history.

In this work we follow the evolution of helium stars with initial
mass $\approx (2.2 - 2.5)\,M_\odot$ and show that they might reach
thermonuclear explosion and perhaps account for some of the peculiar
SNe.

\section{Results}

We followed in detail the evolution of a $2.4\,M_\odot$ helium star,
starting from a homogeneous object. We used the TYCHO evolutionary
code described in \citet{Young_Arnett2005ApJ}. The helium burning
convective core has an almost constant mass of about
$0.82\,M_\odot$, and produces a CO core with $X_{\rm C} \approx
0.27$. Subsequently, a radiative helium burning shell develops above
the core, and as the core grows to about $1.1\,M_\odot$ carbon is
ignited in the core at about $0.3\,M_\odot$ (Fig.~1, left panel).
After the end of core helium burning the center of the star
becomes increasingly degenerate, and neutrino cooling is
increasingly competing with contraction-induced heating. As
a result, the maximum temperature is attained at an off-center point
and it keeps growing until carbon ignites there (Fig.~1, right
panel).
Following off-center C-ignition, the center expands and cools, and a
series of carbon burning flashes occurs. Eventually burning ceases
when carbon is almost exhausted in the core, however as can be seen
in Fig. 2, the innermost $0.5\,M_\odot$ of the core has a residual
$X_{\rm C} \approx 0.02$.

\begin{figure}[!ht]
  \label{fig:hr}
  \plotfiddle{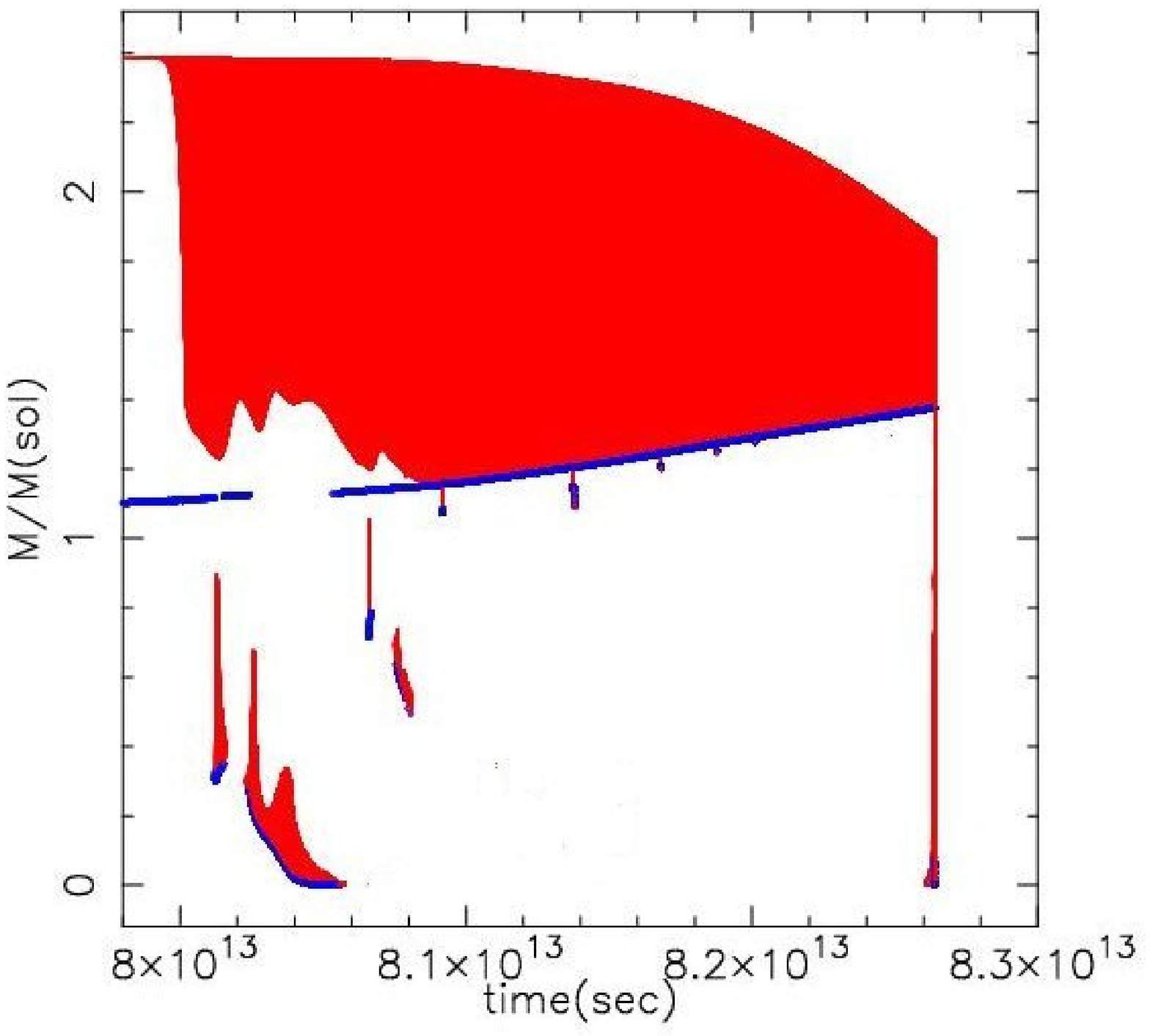}{140pt}{0}{30}{30}{-190}{0}
  \plotfiddle{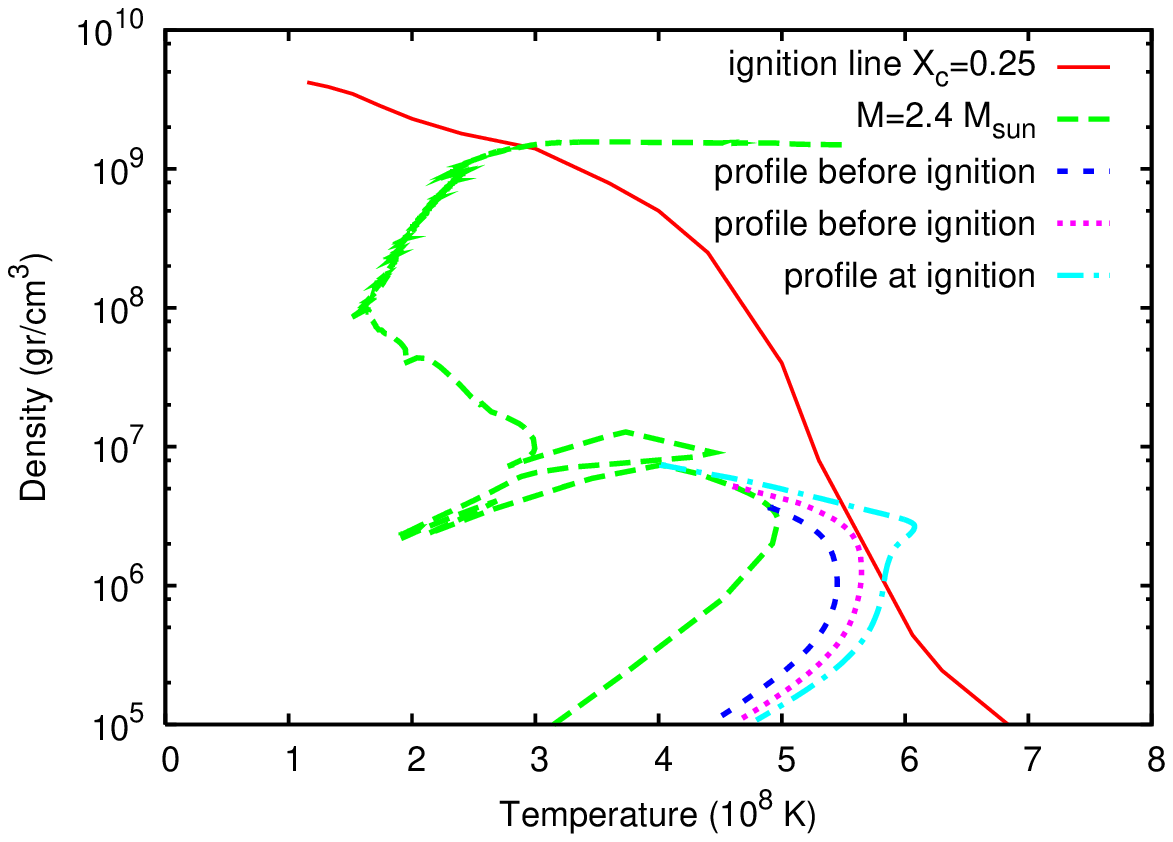}{0pt}{0}{60}{68}{-60}{-20}
  \caption{Evolution of the initially $2.4\,M_\odot$ helium star model.
   \emph{Left:} Kippenhahn plot after the end of core helium burning.
   Light (red) filled areas are convective, dark (blue) line shows maximum thermonuclear
   energy production rate.
   \emph{Right:} Evolution of stellar center on temperature vs. density plane (long dash (green) line);
   also shown is the $\rho-T$ relation of the inner part of the model prior
   (short dashed (dark blue) and dotted (purple) lines)
   and immediately after off-center carbon ignition (dot-dashed (light blue) line).
   Carbon ignition line for $X_{\rm C} = 0.25$ is shown as solid (red) line.}
\end{figure}

After carbon burning ceases, the center continues to heat and
contract while the helium burning shell gradually increases the mass
of the core. Similarly to AGB evolution, the luminosity of the star
grows, causing the envelope to expand and to develop a deepening
convective region, which at certain moment penetrates into the
helium burning shell and creates conditions for hot bottom burning.
During this stage the carbon that accumulates below the helium
burning shell quasi-periodically ignites (see Fig. 1, left panel),
similarly to the helium flashes occurring in AGB stars. Since very
little is known about the mass loss rate of stars of the kind we
consider, we applied \citet{Reimers1975MSRSL}-based rate, which
reduced the mass of the star to $\simeq 1.8\,M_{\odot}$ (Fig.~1,
left panel).

\begin{figure}[!ht]
  \label{figcomp}
  \begin{center}
   \includegraphics[angle=-90,scale=0.4]{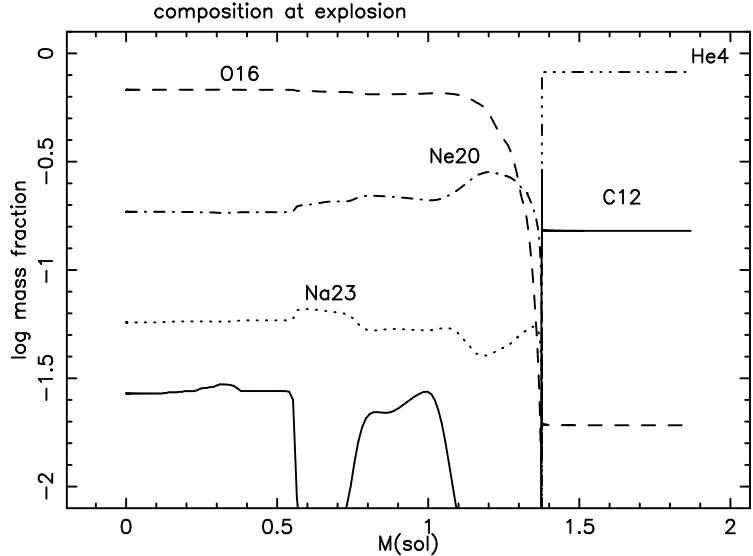}
       \caption{Composition of the initially $2.4\,M_\odot$ helium star model before thermonuclear
  runaway.}
  \end{center}
\end{figure}

Finally, the core grows to $M_{Ch}$ due to the helium burning shell
and carbon in the center ignites (Fig. 1, right panel).
Convection is initiated, supplying more carbon to the central
region, and if the amount of carbon is sufficient to raise the
temperature above the oxygen ignition threshold ($\simeq 1.5 \times
10^9$\,K), which indeed happens in this case, oxygen will ignite
under degenerate conditions and initiate an explosion similar to the
classical SN~Ia case.

Models with initial masses $(2.2 - 2.5)\,M_\odot$ evolve very
similarly to the $2.4\,M_\odot$ case, developing a carbon residue
which might ignite thermonuclear runaway. We did not follow the
evolution further through the hot-bottom burning stage. The
$2.1\,M_\odot$ model only ignites carbon at a later stage, below the
helium burning shell, similar to the carbon flashes encountered in
the $2.4\,M_\odot$ case, however, since in this case carbon
has not been previously depleted in the core, burning will probably
continue until it reaches the center. We have not followed this
model further. The
models of $2.6\,M_\odot$ and more massive ones ignite carbon very
close to or at the center, so that the carbon residue is
either non-existent or insufficient for thermonuclear runaway.

In an earlier work \citep{Waldman2007} we explored the similar case
of CO white dwarfs undergoing off-center carbon burning followed by
mass accretion. As a function of the amount of residual carbon
 we got thermonuclear runaway at a broad
range of $\rho_c \approx (1 - 6) \times 10^9 \mathrm{g\,cm^{-3}}$.
Since the structure of the CO core in our helium star models is very
similar to that of the mass accreting CO white dwarfs, we expect to
get a similar result at runaway.

Supernova which results from thermonuclear runaway in the remnant of
He-star most probably will not differ photometrically from a
``normal'' SN~Ia, but one may expect presence of strong He-lines in
the spectrum, thanks to thick He-mantle of pre-SN, making this SN~Ia
``peculiar'' (N. Chugai, \textit{priv. comm.}).

To complete the picture, we tested whether a hydrogen-deficient star
very similar to our initial models could be created as a result of
close binary evolution. We begun with a binary of $(8.5 +
7.7)\,M_\odot$, with an orbital period of 150 day. After the stage
of Roche lobe overflow, the primary has a total mass of
$2.3\,M_\odot$, a CO core of $1.2\,M_\odot$, and a surface hydrogen
mass fraction of $0.14$. Later, wind mass loss and sporadic RLOF
reduce the mass to $2.1\,M_{\odot}$. Further evolution of the
remnant is similar to the above described.

\vskip 0.4cm

\noindent{To summarize, we showed that evolution of helium stars
with initial mass about $(2.2 - 2.5)\,M_\odot$, which might be
produced in close binaries, may suggest a SN scenario in which
thermonuclear runaway in $M_{Ch}$-mass cores is initiated by a very
small amount of residual carbon, while the stars still have a
several $0.1\,M_{\odot}$\ envelope consisting mostly of helium,
carbon and possibly some hydrogen. In order to give a well justified
statement on the observational outcome, our pre-SN models should be
used for detailed simulations of the explosive runaway and spectra
modeling.

\acknowledgements 
This research was supported in part by the National Science
Foundation under grant PHY05-51164, Russian Academy of Sciences
Basic Research Program ``Origin and Evolution of Stars and Galaxies''
and by Russian Foundation for Basic Research grant 07-02-00454. LRY
acknowledges N.N. Chugai for fruitful discussions and P.P.
Eggleton for providing his evolutionary code and advise on its
usage. RW acknowledges D. Arnett for providing his TYCHO
evolutionary code for public use.


\end{document}